\documentclass[aps,prl,twocolumn]{revtex4}
\usepackage{amssymb}
\usepackage{graphicx}
\usepackage{epsfig}

\begin{document}

\date{\today}
\title{A Simple Model for ``Simple Aging'' in Glassy Yttrium-Hydrides $YH_{3-\delta}$}
\author{M.\ M\"uller}
\affiliation{Department of Physics, Rutgers University, Piscataway, New Jersey 08854 }

\begin{abstract}
A simple explanation for the logarithmic aging of the photoconductivity in $YH_{3-\delta}$ is proposed. We show that the scaling (``simple'' aging) of the relaxation response with the illumination time $t_w$ is consistent with the superposition of independently relaxing excitations with time offsets distributed over a window of width $t_w$. 
\end{abstract}

\maketitle

In a recent article~\cite{LeeRosenbaum05}, M.~Lee et al. reported interesting glassy behaviour in the relaxation of photo-induced conductivity 
in doped yttrium hydrides, $YH_{3-\delta}$ with $\delta\approx 0.039$. The relative decrease of the conductivity was found to be a scaling function of the ratio between the time $t$ elapsed after the excitation, and the illumination time $t_w$, $\delta\sigma(t;t_{w})/\Delta\sigma_{tot}(t_{w})={\cal F}(t/t_w)$, where $\Delta\sigma_{tot}(t_{w})$ is the total excess conductivity induced by the illumination.
This scaling is very similar to the ``full'' or ``simple'' aging observed in polymers, spin glasses~\cite{BouchaudCugliandolo98} and electron glasses~\cite{vaknin00}, except for the fact that the normalization  with the amplitude $\Delta\sigma_{tot}(t_w)$ has no counterpart in those systems. 

The observation of full aging is often taken as an indicator for a glass phase with a rugged energy landscape due to strongly frustrated interactions. Indeed, if the energy landscape is pictured as a collection of valleys with a wide distribution of escape times (with a non-integrable tail), full aging generically arises because the typical relaxation time, i.e., the escape time of the last visited valley, is of the order of the time during which the system has explored the phase space~\cite{Bouchaud92}.

However, as we will argue below, in $YH_{3-\delta}$ the scaling of the relaxation function is likely due to a different mechanism which only requires a large distribution of relaxation times for a collection of independent non-interacting excitations, similarly to two-level systems in strong glass formers. This illustrates that the logical conclusion ``full aging $\rightarrow$ interacting, collectively glassy system'' does not always hold. 

We propose a simple model to explain the observations of~\cite{LeeRosenbaum05}. The illumination of $YH_{3-\delta}$ with energetic photons creates a number of local excitations, each of which independently contributes to the increase of hopping carriers. The precise nature of these excitations is not known. Possible mechanisms could be changes in the bonding configuration between $Y$ and $H$ or the creation of pairs of close hydrogen vacancies by the light-induced hopping of vacancies.
Below the saturation threshold, the increase of conductivity is roughly proportional to the number of excitations, and thus proportional to the total photon energy injected ($\Delta\sigma_{tot}(t_{w})\sim t_w$).

The excitations relax very slowly, presumably via tunneling processes, as suggested by the temperature independence of the experimental data (for $T<140K$). If we assume a broad distribution of tunneling barriers the number of excitations decreases logarithmically with time.
More precisely, the fraction $f(t,t')$ of excitations which were created at time $t'$ and have relaxed by the time $t$, grows as $f(t,t')=C \log[(t-t')/t_0]$ where $C\approx 1/\log(t_M/t_0)$ and $t_0, t_M$ are the shortest and the longest relaxation times, respectively.

After illumination over a time window $t_w$ and an additional relaxation period $t$, the number of remaining excitations, and hence the excess photoconductivity, is proportional to
\begin{equation} 
\sigma(t;t_w)\approx \Delta\sigma_{tot}(t_w)\int_{-tw}^0 (1-C \log[(t-t')/t_0]) \frac{dt'}{t_w}.
\end{equation}
For the relative decrease of the conductivity after illumination we obtain the result 
\begin{eqnarray} 
\label{scalingform}
\frac{\sigma(0;t_w)-\sigma(t;t_w)}{\sigma(0;t_w)}= C\int_0^{t_w} \frac{dt'}{t_w} \log\left[\frac{t+t'}{t'}\right] \\
\,\,\quad= C\left[(t/t_w+1)\log(t/t_w+1) -t/t_w\log(t/t_w)\right],\nonumber
\end{eqnarray}
which fits very well~\cite{Minhyeaprivate} the scaling function found in Ref.~\cite{LeeRosenbaum05}. The good agreement with this simple model suggests that the glassiness in $YH_{3-\delta}$ is of a similar type as that of strong glass formers. 
Note that the necessity to normalize the relaxation by the total excess conductivity $\Delta\sigma_{tot}(t_w)$ is very natural within this model.
The validity of this scenario, and in particular, the correlations between excitations could be tested by studying the conductance noise. According to our simple model we expect it to be very similar to that in strong glass formers with independent two-level systems.   

We note an interesting property of the logarithmic relaxation kernel which distinguishes it from other time-translational invariant kernels, $f(t,t')\equiv f(t-t')$. One can prove that the relative conductivity decrease takes a scaling form ${\cal F}(t/t_w)$ {\em if and only if} $f(\tau)= C\log(\tau/t_0)$, in which case ${\cal F}(t/t_w)$ is necessarily of the universal form (\ref{scalingform}). 

In the above model the ``full'' aging (\ref{scalingform}) arises simply due to the superposition of logarithmic relaxations with different temporal offsets. The scaling of the relaxation data therefore does not imply the existence of a rugged free energy landscape or a glass phase due to strong electron-electron interactions.
However, the scenario of independently relaxing excitations can certainly not apply in a situation where the observed scaling function ${\cal F}$ deviates significantly from Eq.~(\ref{scalingform}). This seems to be the case the for aging experiments in indium-oxides~\cite{vaknin00}, where a logarithmic behaviour is observed only for times \textit{shorter} than $t_w$, and a crossover to faster relaxation is observed around $t_w$. In contrast, the scaling function (\ref{scalingform}) assumes a logarithmic behaviour only for times $t\geq t_w$, without saturating on experimental time scales in $YH$.
 
We thank M. Lee and T. Rosenbaum for discussions.
\vspace{-0.45cm}

\bibliographystyle{prsty}

\begin{thebibliography}{1}

\bibitem{LeeRosenbaum05}
M. Lee {\it et~al.}, J. Phys. Condens. Matter {\bf 17},  L439  (2005).

\bibitem{BouchaudCugliandolo98}
J.-P. Bouchaud, L. Cugliandolo, J. Kurchan, and M. M{\'e}zard,  in {\em Spin
  Glasses and Random Fields}, edited by A.~P. Young (World Scientific,
  Singapore, 1998).

\bibitem{vaknin00}
A. Vaknin, Z. Ovadyahu, and M. Pollak, Phys. Rev. Lett. {\bf 84},  3402
  (2000). V. Orlyanchik and Z. Ovadyahu, Phys. Rev. Lett. {\bf 92},  066801  (2004).

\bibitem{Bouchaud92}
J.~P. Bouchaud, J. Phys. I France {\bf 2},  1705  (1992).

\bibitem{Minhyeaprivate}
M. Lee private communication.

\end{thebibliography}

\end{document}